\documentclass[showpacs, twocolumn]{revtex4-1}
\usepackage{graphicx}
\usepackage{amsmath}
\usepackage{appendix}

\begin{document}

\title{Dipolar Bose gas in a weak isotropic speckle disorder}

\author{Abdel\^{a}ali Boudjem\^{a}a}

\affiliation{Department of Physics, Faculty of Sciences, Hassiba Benbouali University of Chlef P.O. Box 151, 02000, Ouled Fares, Chlef, Algeria.}


\begin{abstract}
We investigate the properties of a homogeneous dipolar Bose gas in a weak three-dimensional (3D) isotropic speckle disorder at finite temperature. 
By using the Bogoliubov theory (beyond the mean field), we calculate the condensate and the superfluid 
fractions as a function of density  and strengths of disorder and interaction.
The influence of disorder on the anomalous density, the chemical potential and the ground state energy is also analyzed.
We show that the peculiar interplay of the DDI and weak disorder makes the superfluid fraction and sound velocity anisotropic.
\end{abstract}


\maketitle

\section{Introduction}

Disordered Bose gas in a weak random external potential (dirty boson) represents an interesting model for studying the relation between Bose-Einstein condensation (BEC)
and superfluidity and has been the subject of many theoretical investigations in the last two decades \cite {Lee, Huang, Gior, Mich, Lopa, Gior1, Axel, Yuk,Lug}.
Experimentally, the dirty boson problem, was first studied with superfluid helium in aerosol glasses (Vycor)\cite {Crook, Fish, Chan}. 
Recently, several groups have loaded ultracold atoms into optical potentials and studied BECs in the presence of disorder \cite {Schul, Sanch, Asp, Asp1,White, Dries}.

What happens to a homogeneous BEC if a weak random external potential is switched on? 
Indeed, the presence of disordered potential may lead to decrease both BEC and superfluidity.
Furthermore, one of the intriguing feature of disordered  Bose gas is the appearance of the so-called Anderson localization \cite {And} in the non-interacting case.
This phenomenon which can be understood as the effect of multiple reflections of a plane wave by random scatterers or random potential barriers, 
has recently attracted a great deal of interest \cite {Sanch, Asp, Asp1}.
Experimentally, the random potential can be created using different techniques, one of which is the static laser speckle, 
whereas potential felt by atoms is proportional to the speckle intensity with the sign of the detuning from the atomic transition \cite{Clem1}. 
Laser speckle, produced by passing expanded laser beam through diffusive plates, are special in that they have (i) exponential, i.e. strongly non-Gaussian, intensity distribution 
and (ii) finite support of their power spectrum \cite{Godm}. 
Recent progress in different experimental realizations of laser speckle disorder is reported in references \cite {Clem1, Kond}.

In their recent work Abdulaev et\textit {al} \cite {Abdulaev}, have shown that a Gaussian approximation of the autocorrelation function of laser speckles, used
in some recent papers, is inconsistent with the general background of laser speckle theory. 
They also pointed out that the concept of a quasi-3D speckle, which appears due to an extension
of the autocorrelation function in the longitudinal direction of a transverse 2D speckle, is not applicable
for the true 3D speckle, since it requires an additional space dimension. 
In this context, they derived an appropriate autocorrelation function for an isotropic 3D laser speckle potential which has the Fourier transform given in Eq.(\ref{CorrS}) (see below).

Recent progress in the physics of ultra-cold gases have led to the creation of BECs with dipole-dipole interaction (DDI ) and stimulated a tremendous boost in theoretical 
and experimental studies of weakly interacting Bose gases \cite {Baranov, Pfau, Pupillo2012}. 
What is important in such systems is that the atoms interact via a DDI 
that is both long ranged and anisotropic. By virtue of this interaction, these systems are expected to open fascinating prospects for the observation of novel quantum phases
in ultracold atomic gases. On the other hand, dipolar BECs confined in random media remain largely unexplored. One can quote for example, 
uniform dipolar Bose gas with a Gaussian disorder correlation function, a Lorentzian, and a delta-correlated disorder have been explored recently by Pelster et\textit {al}
\cite {Axel1, Axel2, Axel3}.

In the present paper, we study the impact of a weak disorder potential with 3D isotropic laser speckle autocorrelation function of Ref.\cite {Abdulaev}
on the properties of a homogeneous dipolar Bose gas at finite temperature.
To this end, we use the Bogoliubov theory (beyond the mean field) and we 
calculate in particular the condensed depletion and the anomalous fraction. This latter quantity which grows with increasing interactions and vanishes
in noninteracting systems\cite {Griffin,  boudj2011, boudj2012, boudj2015}, is important to fully understand the interplay of disorder and interactions.
We show, in addition, how the anisotropy of the DDI enhance quantum, thermal and disorder fluctuations as well as the superfluid fraction.

The rest of the paper is organized as follows. In Sec.\ref {model}, we describe our model of the dipolar dilute Bose gas in a general random potential.
In Sec.\ref{zeroT}, we derive analytical expressions for the condensate fluctuations and some 
thermodynamic quantities for 3D isotropic laser speckle disorder potential at finite temperature.
We show that the competition between both contact interaction-disorder and DDI-disorder leads to enhance the condensate depletion, 
the anomalous density, disorder fluctuation, ground state energy, equation of state and the sound velocity. 
In Sec.\ref{superfluid}, the superfluid fraction is obtained and its characteristics are discussed.
Finally, our conclusions and outlook remain in Sec.\ref{conc}.

\section{The model} 
\label {model}
We consider the effects of an external random field on a  dilute 3D dipolar Bose gas with dipoles oriented perpendicularly to the plane.
The Hamiltonian of the system is written as:
\begin{align}\label{ham}
&\hat H = \int d^3r \, \hat \psi^\dagger ({\bf r}) \left(\frac{-\hbar^2 }{2m}\Delta+U({\bf r})\right)\hat\psi(\mathbf{r}) \nonumber \\
&+\frac{1}{2}\int d^3r\int d^3r^\prime\, \hat\psi^\dagger(\mathbf{r}) \hat\psi^\dagger (\mathbf{r^\prime}) V(\mathbf{r}-\mathbf{r^\prime})\hat\psi(\mathbf{r^\prime}) \hat\psi(\mathbf{r}) ,
\end{align}
where $\psi^\dagger$ and $\psi$ denote, respectevly the usual creation and annihilation field operators, the interaction potential
$V(\mathbf{r}-\mathbf{r^\prime})=g\delta(\mathbf{r}-\mathbf{r^\prime})+V_{dd}(\mathbf{r}-\mathbf{r^\prime})$, $g=4\pi \hbar^2 a/m$
corresponds to the short-range part of the interaction and is parametrized by the scattering length as $a$. On the other hand, the dipole-dipole component reads  
\begin{equation}\label{dd}
V_d(\vec r) =\frac{C_{dd}}{4\pi}\frac{1-3\cos^2\theta}{r^3},
\end{equation}
where the coupling constant $C_{dd} $ is $M_0 M^2$ for particles having a permanent magnetic dipole moment $M$ ($M_0$ is the magnetic permeability
in vacuum) and $d^2/\epsilon_0$ for particles having a permanent electric dipole $d$ ($\epsilon_0 $ is the permittivity of vacuum),
$m$ is the particle mass, and $\theta$ is the angle between the relative position of the particles $\vec r$ and the direction of the dipole.
The characteristic dipole-dipole distance can be defined as $r_*=m C_{dd}/4\pi \hbar^2$. For most polar molecules $r_*$ ranges from 10 to $10^4$ \AA. 
The disorder potential is described by vanishing ensemble averages $\langle U(\mathbf r)\rangle=0$
and a finite correlation of the form $\langle U(\mathbf r) U(\mathbf r')\rangle=R (\mathbf r,\mathbf r')$.

Passing to the Fourier transform and working in the momentum space, the Hamiltonian (\ref {ham}) takes the form:
\begin{align}\label{ham1}
&\hat H\!\!=\!\!\sum_{\bf k}\! \frac{\hbar^2k^2}{2m}\hat a^\dagger_{\bf k}\hat a_{\bf k}\! +\!\frac{1}{V}\!\!\sum_{\bf k,\bf p} \! U_{\bf k\!-\!\bf p} \hat a^\dagger_{\bf k} \hat a_{\bf p}  \\ \nonumber
&+\!\frac{1}{2V}\!\!\sum_{\bf k,\bf q,\bf p}\!\!
f ({\bf p})\hat a^\dagger_{\bf k\!+\!\bf q} \hat a^\dagger_{\bf k\!-\!\bf q}\hat a_{\bf k\!+\!\bf p}\hat a_{\bf k\!-\!\bf p} ,
\end{align}
where $V$ is a quantization volume, and the interaction potential in momentum space is given by \cite{boudj2015}
\begin{equation}\label{scam}
 f(\mathbf k)=g[1+\epsilon_{dd} (3\cos^2\theta-1)], 
\end{equation}

Assuming the weakly interacting regime where  $r_*\ll \xi$ with $\xi=\hbar/\sqrt{mgn}$ being the healing length and $n$ is the total density,
we may use the Bogoliubov approach. Applying the inhomogeneous Bogoliubov transformations \cite{Huang}: 
\begin{equation}\label {trans}
 \hat a_{\bf k}= u_k \hat b_{\bf k}-v_k \hat b^\dagger_{-\bf k}-\beta_{\bf k},  \qquad \hat a^\dagger_{\bf k}= u_k \hat b^\dagger_{\bf k}-v_k \hat b_{-\bf k}-\beta_{\bf k}^*,
\end{equation}
where $\hat b^\dagger_{\bf k}$ and $\hat b_{\bf k}$ are operators of elementary excitations.
The Bogoliubov functions $ u_k,v_k$ are expressed in a standard way:
$ u_k,v_k=(\sqrt{\varepsilon_k/E_k}\pm\sqrt{E_k/\varepsilon_k})/2$ with $E_k=\hbar^2k^2/2m$ is the energy of a free particle, and 
\begin{equation}\label {beta}
\beta_{\bf k}=\sqrt{\frac{n}{V}} \frac{E_k}{\varepsilon_k^2}  U_k .
\end{equation}
The Bogoliubov excitations energy is given by 
\begin{equation}\label {spec}
\varepsilon_k=\sqrt{E_k^2+ 2\mu_{0d}(\theta) E_k},
\end{equation}
where $\mu_{0d}=n\lim\limits_{k\rightarrow 0} f({\bf k})$ is the zeroth order chemical potential.\\
Importantaly, the spectrum (\ref {spec}) is independent of the random potential. 
This independence holds in fact only in zeroth order in perturbation theory; conversely, higher order calculations render the spectrum dependent on the random potential
due to the contribution of the anomalous terms (see below).
For $k\rightarrow 0$, the excitations are sound waves $\varepsilon_k=\hbar c_{sd} (\theta) k$, where $c_{sd} (\theta)=c_{s}\sqrt{1+\epsilon_{dd} (3\cos^2\theta-1)}$
with $c_{s}=\sqrt{gn/m}$ is the sound velocity without DDI.
Due to the anisotropy of the dipolar interaction, the sound velocity acquires a dependence on the propagation direction, which is fixed by the angle
$\theta$ between the propagation direction and the dipolar orientation. This angular dependence of the sound velocity has been confirmed experimentally \cite {bism}.

Therefore, the diagonal form of the Hamiltonian of the dirty dipolar Bose gas (\ref{ham1}) can be written as
\begin{equation}\label{DHami} 
\hat H = E+\sum\limits_{\vec k} \varepsilon_k\hat b^\dagger_{\bf k}\hat b_{\bf k},
\end{equation}
where $E=E_{0d}+\delta E+ E_R$, \\ $E_{0d}(\theta)=\mu_{0d}(\theta) N /2$ with $N$ being the total number of particles. 
\begin{equation}\label{Fenergy} 
\delta E=\frac{1}{2}\sum\limits_{\bf k} [\varepsilon_k -E_k-n f({\bf k})],
\end{equation}
is the ground-state energy correction due to qunatum fluctuations. 
\begin{equation}\label{Renergy} 
E_R=-\sum\limits_{\bf k} n\langle |U_k|^2\rangle \frac{ E_k}{\varepsilon_k^2} =-\sum\limits_{\bf k} n R_k \frac{ E_k}{\varepsilon_k^2},
\end{equation}
gives the correction to the ground-state energy due to the external random potential.

The noncondensed and the anomalous densities are  defined as  $\tilde{n}=\sum_{\bf k} \langle\hat a^\dagger_{\bf k}\hat a_{\bf k}\rangle$ 
and $\tilde{m}=\sum_{\bf k} \langle\hat a_{\bf k}\hat a_{-\bf k}\rangle$, respectively. Then invoking for the operators $a_k$
the transformation (\ref {trans}), setting $\langle \hat b^\dagger_{\bf k}\hat b_{\bf k}\rangle=\delta_{\bf k' \bf k}N_k$ and putting the rest of the expectation values equal to zero,
where $N_k=[\exp(\varepsilon_k/T)-1]^{-1}$ are occupation numbers for the excitations.  
As we work in the thermodynamic limit, the sum over $k$ can be replaced by the integral $\sum_{\bf k}=V\int d^3k/(2\pi)^3$ 
and using the fact that $2N (x)+1= \coth (x/2)$, we obtain:
\begin{equation}\label {nor}
\tilde{n}=\frac{1}{2}\int \frac{d^3k} {(2\pi)^3} \frac{E_k+f({\bf k}) n} {\varepsilon_k}\left[\coth\left(\frac{\varepsilon_k}{2T}\right)-1\right]+n_R,
\end{equation}
and
\begin{equation}\label {anom}
\tilde{m}=-\frac{1}{2}\int \frac{d^3k} {(2\pi)^3} \frac{f({\bf k}) n} {\varepsilon_k}\coth\left(\frac{\varepsilon_k}{2T}\right)+n_R.
\end{equation}
The contribution of the random potential comes through the last terms in Eqs (\ref{nor}) and (\ref {anom}). These terms are defined as 
\begin{equation}\label{depdis}
n_R=\frac{1}{V}\sum\limits_{\bf k} \langle |\beta_{\bf k}|^2\rangle=n\int \frac{d^3k} {(2\pi)^3} \frac{ E_k^2}{\varepsilon_k^4} R_k .
\end{equation}

Expressions (\ref{nor}) and (\ref {anom}) must satisfy the equality 
\begin{align}\label {heis}
\tilde{n}_k(\tilde{n}_k+1)-|\tilde{m}_k|^2&= \frac{1}{4\, \text {sinh}^2\left(\varepsilon_k/2T\right)} \nonumber\\
&+n_R\left(\frac{E_k+2f({\bf k})n}{\varepsilon_k} \right) \text {coth} \left(\frac{\varepsilon_k}{2T}\right).
\end{align}
Equation (\ref {heis}) clearly shows that $\tilde{m}$ is larger than $\tilde{n}$ at low temperature irrespective to the presence of an external random potential or not. 
So the omission of the anomalous density in this situation 
is principally unjustified approximation and wrong from the mathematical point of view \cite{boudj2011, boudj2012, boudj2015}.

\section{BEC fluctuations and thermodynamics quantities in weak isotropic laser speckle}
\label {zeroT}
To proceed further in practical calculations, we must define the laser speckle potential:  
$U({\bf r})=U_0+\Delta U({\bf r})$, where $U_0$ is defined by the light far-field intensity as $U_0=\langle I\rangle$ and $\langle \Delta U({\bf r})\rangle=0$.
At the derivation of $U({\bf r})$, it was assumed that the incident laser wave does not induce
an atomic electron interlevel transition, but merely deforms the atomic ground state.
It is useful now to specify the relationship between the far-field intensity autocorrelation function 
$|C_I({\bf r})|^2$, the laser speckle autocorrelation function $|C_A({\bf r})|^2$ and the disorder potential correlation function. One can write then:
$|C_I({\bf r})|^2=\langle U({\bf r'}) U({\bf r'+r})\rangle$ and $|C_A({\bf r})|^2=\langle \Delta U({\bf r'}) \Delta U({\bf r'+r})\rangle/U_0^2$.
Therefore, using the Fourier transform, we get
\begin{equation} \label{Disp}
|C_I({\bf k})|^2 = U_0 ^2[\delta({\bf k})+|C_A({\bf k})|^2],
\end{equation}
where the autocorrelation function of the laser speckle is given by \cite {Abdulaev}
\begin{equation} \label{CorrS}
|C_A({\bf k})|^2 = \frac{3}{4\pi} (2\sigma)^3[(2\sigma k)^3-12(2\sigma k)+16],
\end{equation}
where $\sigma$ characterize the correlation length of the disorder (for further computational details, see Ref \cite {Abdulaev}).
Interestingly, we see from the formula of $|C_A({\bf k})|^2$ that its value becomes zero for
$k=1/\sigma$. Hence, the momentum in (\ref{CorrS}) only varies in a finite interval from zero, in contrast to the case for a Gaussian function\cite{Axel1}. 

\begin{figure}[htb1] 
\includegraphics[scale=0.8]{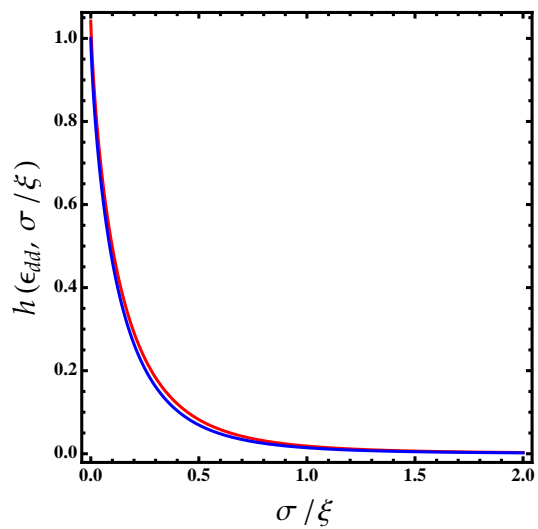}
\caption {(Color online) Behavior of the disorder function $h(\epsilon_{dd}, \sigma/\xi)$ from equation (\ref {func}),
 as a function of $\sigma/\xi$.  Red line: Er atoms ($\epsilon_{dd}=0.38$).  Blue line: pure contact interaction ($\epsilon_{dd}=0$).}
\label{dis}
\end{figure}

Putting $R({\bf k})=R|C_A({\bf k})|^2 $ \cite{Abdulaev}, where $R=U_0^2$ stands for the disorder strength.
Substituting now the function (\ref{CorrS}) in equation (\ref{depdis}) and performing the integration over the momentum form 0 to $1/\sigma$, we get the expression
for the condensate fluctuation due to the external random potential 
\begin{equation} \label{depdis1}
{n_R}=\frac{m^2R}{8\pi^{3/2} \hbar^4} \sqrt{\frac{n}{a}} h(\epsilon_{dd}, \alpha),
\end{equation}
where 
\begin{equation} \label{func}
h(\epsilon_{dd}, \alpha)= \int_0^\pi d\theta\frac{\sin\theta S(\alpha)}{\sqrt{1+\epsilon_{dd} (3\cos^2\theta-1)}}, 
\end{equation}
is depicted in Fig.\ref{dis}, and the function 

\begin{equation} \label{funcS}
 \begin{split}
S(\alpha)=\frac{1}{2\pi}\sqrt{\frac{\alpha}{2}} \left[4-(8\alpha+6)\ln\left(1+\frac{1}{2\alpha}\right)\right. \\+
\left.2\sqrt{\frac{2}{\alpha}}\arctan\left(\frac{1}{\sqrt{2\alpha}}\right)\right]. \nonumber
\end{split}
\end{equation}
 with
$\alpha=\sigma^2[1+\epsilon_{dd} (3\cos^2\theta-1)]/\xi^2$.  \\

In the absence of the DDI ($\epsilon_{dd}=0$), we recover the result for the 3D BEC with short-range interparticle interaction of Ref \cite {Abdulaev}.
For $\sigma/\xi\rightarrow 0$ and $\epsilon_{dd}=0$, we read off from Eq.(\ref{func}) that one obtains $h(\epsilon_{dd}, \alpha) \rightarrow 1$ (see also Fig.\ref{dis}). 
Therefore, we should reproduce the Huang and Meng result \cite{Huang} for the condensate depletion in this limit. 
For $\sigma/\xi\rightarrow 0$, we get from Eq.(\ref{func}) that $h(\epsilon_{dd}, 0) ={\cal Q}_{-1}(\epsilon_{dd})$. 
Thus, the disorder fluctuation (\ref{depdis1}) becomes identical to that obtained in 3D dipolar BEC with delta correlated disorder \cite {Axel3} 
\begin{equation} \label{depdis2}
{n_R}=\frac{m^2R}{8\pi^{3/2} \hbar^4} \sqrt{\frac{n}{a}} {\cal Q}_{-1}(\epsilon_{dd}),
\end{equation}
where the contribution of the DDI is expressed by the functions ${\cal Q}_j(\epsilon_{dd})=(1-\epsilon_{dd})^{j/2} {}_2\!F_1\left(-\frac{j}{2},\frac{1}{2};\frac{3}{2};\frac{3\epsilon_{dd}}{\epsilon_{dd}-1}\right)$, where ${}_2\!F_1$ is the hypergeometric function. Note that functions ${\cal Q}_j(\epsilon_{dd})$ attain their maximal values for $\epsilon_{dd} \approx 1$ 
and become imaginary for $\epsilon_{dd}>1$ \cite {lime, boudj2015}.\\
On the other hand, the function (\ref {func}) decreases with increasing disorder correlation length while it rises for increasing $\epsilon_{dd}$ and diverges in the limit $\epsilon_{dd}>1$. 
Another important consequence is that when $a$ vanishes, $n_R$ becomes infinite. This means that the system would collapse if there were no repulsive interactions between particles. 

Upon calculating integral in Eq.(\ref{nor}), we  get for the condensate depletion 
\begin{align}\label {dep}
\frac{\tilde{n}}{n}=&\frac{ 8}{3} \sqrt{\frac{ n a^3}{\pi}} {\cal Q}_3(\epsilon_{dd})+\frac{2}{3}\sqrt{\frac{n a^3}{\pi}} \left(\frac{\pi T}{gn}\right)^{2}{\cal Q}_{-1}(\epsilon_{dd}) \nonumber\\
&+2\pi R' \sqrt{\frac{ n a^3}{\pi}} h(\epsilon_{dd}, \alpha),
\end{align}
where $R'=R/g^2n$ is a dimensionless disorder strength.

The integral in Eq.(\ref{anom}) is ultraviolet divergent. This divergence is well-known to be unphysical, since it is caused by
the usage of the contact interaction potential. A general way of treating such integrals is as follows. First, one restricts to
asymptotically weak coupling and introduces the Beliaev-type second order coupling constant \cite {boudj2015}
\begin{equation}\label {cc}
f_R ({\bf k})=f({\bf k}) -\frac{m}{\hbar^2}\int \frac {d^3q}{(2\pi)^3} \frac{f(-{\bf q}) f({\bf q})}{2E_q}.
\end{equation}
After the subtraction of the ultraviolet divergent part, the anomalous fraction turns out to be given 
\begin{align}\label {anom1}
\frac{\tilde{m}}{n}=&8\sqrt{\frac{ n a^3}{\pi}} {\cal Q}_3(\epsilon_{dd})-\frac{2}{3}\sqrt{\frac{n a^3}{\pi}} \left(\frac{\pi T}{gn}\right)^{2}{\cal Q}_{-1}(\epsilon_{dd})\nonumber\\
&+2\pi R' \sqrt{\frac{ n a^3}{\pi}} h(\epsilon_{dd}, \alpha).
\end{align}
The leading term in Eqs.(\ref{dep}) and (\ref{anom1}) represents the qunatum fluctuation\cite {boudj2015}. 
The subleading term which represents the thermal fluctuation\cite {boudj2015}, is calculated at temperatures $T\ll gn $, 
where the main contribution to integrals (\ref{nor}) and (\ref{anom}) comes from the region of small momenta ($\varepsilon_k=\hbar c_{sd} k$).
The situation is quite different at higher temperatures i.e. $T\gg gn$, where the main contribution to integrals (\ref{nor}) and (\ref{anom}) comes from the single particle excitations. 
Hence, the thermal contribution of $\tilde{n}$ becomes identical to the density of noncondensed atoms in an ideal Bose gas \cite{boudj2015}, while
the thermal contribution of $\tilde{m}$ tends to zero since the gas is completely thermalized in this range of temperature \cite{Griffin, boudj2011, boudj2015}.
The last term in (\ref{dep}) and (\ref{anom1}) accounts for the effect of disorder on the noncondensed and the anomalous densities.

Equation (\ref {anom1}) clearly shows that at zero temperature, the anomalous density is larger than the noncondensed density 
for any range of the dipolar interaction as well as for any value of the strength and the correlation length of the disorder as it has been anticipated above. 
Moreover, $\tilde{m}$ changes its sign with increasing temperature in agreement with uniform Bose gas with a pure contact interaction \cite {boudj2015}.
Likewise, the anomalous density obtained in (\ref {anom1}) permits us to elaborate in a straightforward manner the equation of state and thus, leads to a finite compressibility  (see below).
Remarkably, Eqs.(\ref {dep}) and (\ref{anom1}) reproduce the short range interaction results since ${\cal Q}_j(\epsilon_{dd}=0)=1$.
Furthermore, the DDI enhances quantum, thermal and disorder fluctuations of the condensate for increasing $\epsilon_{dd}$ as is shown in Fig.\ref{dis}.

The Bogoliubov approach assumes that fluctuations should be small. We thus conclude from Eqs. (\ref {dep}) and (\ref {anom1}) that at $T = 0$, the validity of the 
Bogoliubov theory requires inequalities $\sqrt{n a^3}  {\cal Q}_{3} (\epsilon_{dd})\ll 1$ and $ R' \sqrt{n a^3} h(\epsilon_{dd}, \alpha)\ll 1$.
For $R'=0$, this parameter differs only by the factor ${\cal Q}_3(\epsilon_{dd})$ from the universal small parameter of the theory, $\sqrt{na^3}\ll 1$, in the absence of DDI. 
At $T \ll gn$, the Bogoliubov theory requires the condition $(T/gn) \sqrt{n a^3}  {\cal Q}_{-1}(\epsilon_{dd})\ll 1$. The appearance of the extra factor ($T/gn $) 
originates from the thermal fluctuations corrections.

The presence of quantum and disorder fluctuations leads also to corrections of the chemical potential which are given by
$\delta \mu=\sum_{\bf k} f({\bf k}) [v_k(v_k-u_k)]=\sum_{\bf k} f({\bf k}) (\tilde{n}+\tilde{m})$ \cite{boudj2015, boudj2012, Abdougora}. 
Inserting the definitions (\ref{nor}) and (\ref {anom}) into the expression of $\delta \mu$, we find after integration:
\begin{equation}\label {chem}
\frac{\delta\mu} {\mu_0}=\frac{32}{3}\sqrt{\frac{n a^3}{\pi}} {\cal Q}_5(\epsilon_{dd})+4\pi R' \sqrt{\frac{ n a^3}{\pi}} h_1(\epsilon_{dd}, \alpha). 
\end{equation}
where $h_1(\epsilon_{dd}, \alpha)= \int_0^\pi d \theta \sin\theta \sqrt{1+\epsilon_{dd} (3\cos^2\theta-1)} S(\alpha)$
and $\mu_0=gn$.\\
In the absence of disordered potential ($R'=0$), Eq.(\ref{chem}) coincides with that derived recently in \cite {lime, boudj2015}.
For a condensate with a pure contact interaction (${\cal Q}_5(\epsilon_{dd}=0)=1$) and for $R'=0$, 
the obtained correction to the chemical potential (\ref{chem}) excellently agrees with the seminal Lee-Huang-Yang quantum corrected equation of state \cite{LHY}.\\

The energy shift due to the interaction and the quantum fluctuations (\ref {Fenergy})  is ultraviolet divergent. The difficulty is
overcome if one takes into account the second-order correction to the coupling constant (\ref {cc}). A simple calculation yields\cite {lime, boudj2015}
\begin{equation}\label {energ} 
\delta E=\frac{64}{15}Vgn^2\sqrt{\frac{n a^3}{\pi}} {\cal Q}_5(\epsilon_{dd}).
\end{equation} 
However, the energy shift due to the external random potential (\ref {Disp}) is evaluated as 
\begin{equation}\label {Reng}
\frac{E_R}{E_0}=16 \pi R' \sqrt{\frac{n a^3}{\pi}} h_1(\epsilon_{dd}, \alpha),
\end{equation}
where $E_0=N gn/2$.\\
When $\sigma \ll\xi$, the energy shift due to the external random potential (\ref {Renergy}) is ultraviolet divergent. 
Again, by introducing the renormalized coupling constant (\ref{cc}) one gets:
$E_R/E_0=16 \pi R' \sqrt{n a^3/\pi} {\cal Q}_1(\epsilon_{dd})$ which well concides with the result obtained with delta correlated disorder of Ref \cite {Axel3}.


\section{superfluid fraction} \label {superfluid}

\begin{figure}[htb1] 
\centerline{
\includegraphics[scale=0.45]{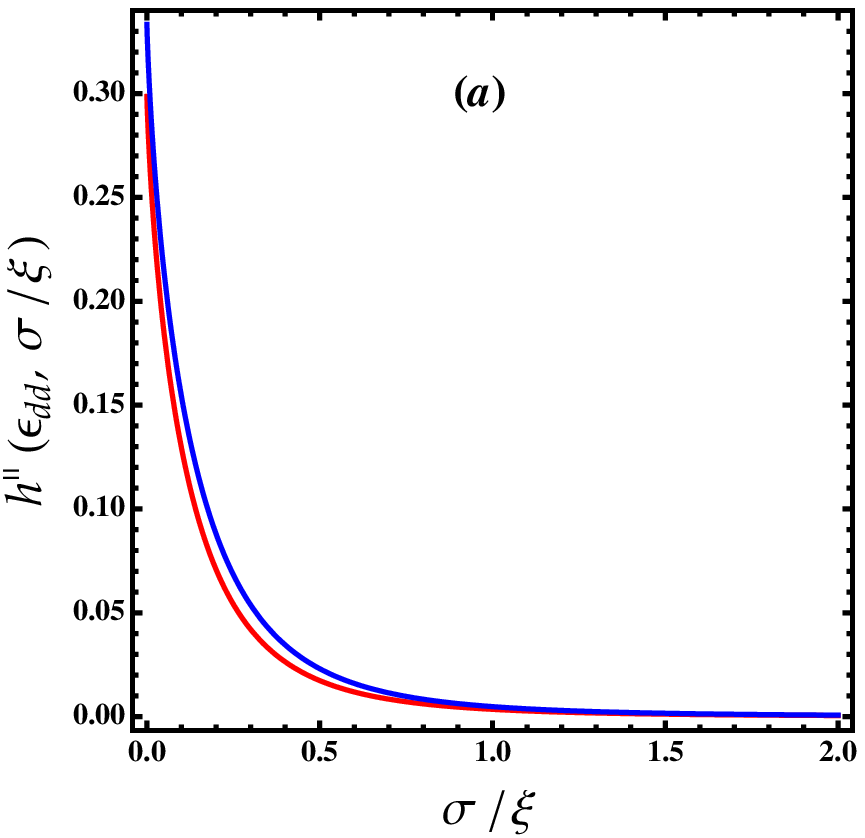}
\includegraphics[scale=0.45]{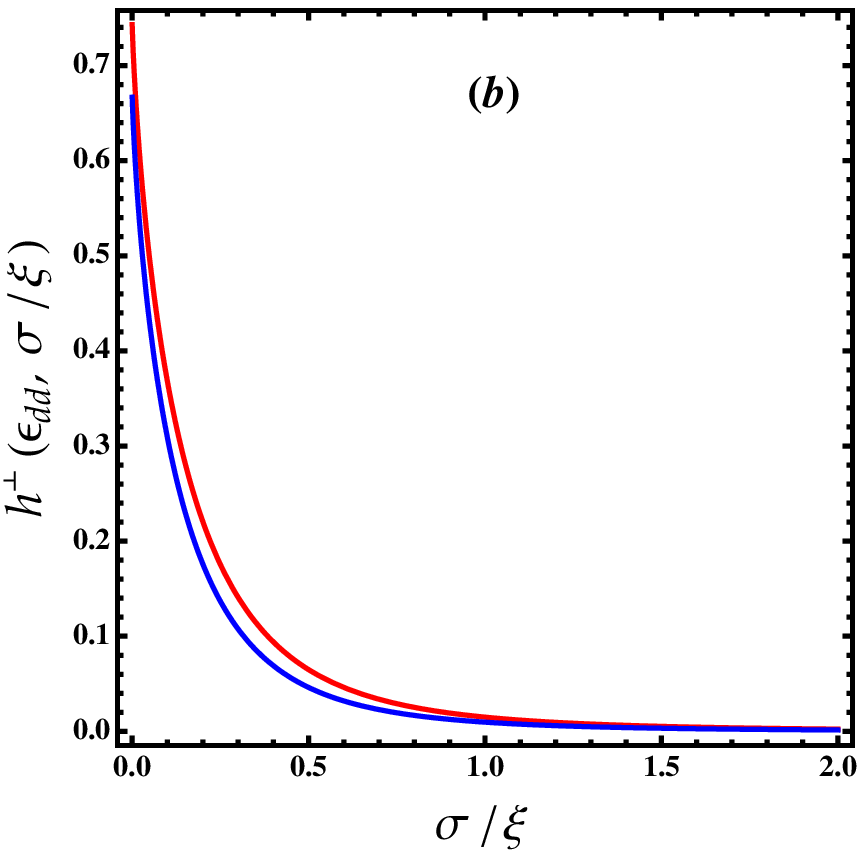}}
\caption {(Color online) Behavior of the disorder functions $h^{\parallel}(\epsilon_{dd}, \sigma/\xi)$ (a) and $h^{\perp}(\epsilon_{dd}, \sigma/\xi)$ (b),
 as a function of $\sigma/\xi$.  Red line: Er atoms ($\epsilon_{dd}=0.38$).  Blue line: pure contact interaction ($\epsilon_{dd}=0$).}
\label{disP}
\end{figure}

The superfluid fraction $n_s/n$ can be found from the normal fraction $n_n/n$ which is determined by the transverse current-current
correlator $n_s/n =1-n_n/n$. We apply a Galilean boost with the total momentum of the moving system ${\bf P}=mv (n {\bf v_s}+n_n {\bf v_n})$, where 
${\bf v_s}$ denotes the superfluid velocity and and ${\bf v_n}={\bf u}-{\bf v_s}$ is normal fluid  velocity with ${\bf u}$ being a boost velocity \cite{Axel2}. 
The superfluid fraction is then written 
\begin{equation} \label{sup1}
 \begin{split}
 \frac{n_s^{ij}}{n}= \delta_{ij}-4\int \frac{d^3k}{(2\pi)^3} \frac{\hbar^2}{2m} \frac{n R_k k_ik_j}{E_k[E_k-2n f({\bf k})]^2} \\ 
-\frac{2}{Tn}\int \frac{d^3k}{(2\pi)^3} \left[\frac{\hbar^2}{2m} \frac{k_i k_j}{4\, \text {sinh}^2 (\varepsilon_k/2T)}\right]. 
\end{split}
\end{equation}
It is worth remarking that if in expression (\ref{sup1}) $\tilde {m}$ were omitted, 
then the related integral would be divergent leading to the meaningless value $n_s \rightarrow−\infty$.
This indicates that the presence of the anomalous density is crucial for the occurrence of the superfluidity in Bose gases \cite{boudj2015, Yuk}
which is in fact natural since both quantities are caused by atomic correlations.\\
Equation. (\ref{sup1})  yields a superfluid density that depends on the direction of the superfluid motion with respect to
the orientation of the dipoles. In the parallel direction, the superfluid fraction reads
\begin{equation}\label {supflui1}
\frac{n_s^{\parallel}}{n} =1-4\pi R' \sqrt{\frac{n a^3}{\pi}} h^{\parallel}(\epsilon_{dd}, \alpha)- \frac{2\pi^2 \hbar}{45 mn c_s} \left(\frac{T}{\hbar c_s}\right)^4 {\cal Q}_{-5}^{\parallel}(\epsilon_{dd}),
\end{equation}
where the function
\begin{equation} \label{funcP}
h^{\parallel}(\epsilon_{dd}, \alpha)= \int_0^\pi d\theta \frac{ \sin\theta\cos^2\theta S(\alpha)}{\sqrt{1+\epsilon_{dd} (3\cos^2\theta-1)}}, 
\end{equation}
is decreasing with increasing $\epsilon_{dd}$ for fixed $\sigma/\xi$ as is depicted in Fig.\ref{disP}.a. And the functions
${\cal Q}_j^{\parallel} (\epsilon_{dd})=\frac{1}{3}(1-\epsilon_{dd})^{j/2} {}_2\!F_1\left(-\frac{j}{2},\frac{5}{2};\frac{3}{2};\frac{3\epsilon_{dd}}{\epsilon_{dd}-1}\right)$, 
have the following properties: ${\cal Q}_j^{\parallel} (\epsilon_{dd}=0)=1/3$ and imaginary for $\epsilon_{dd}>1$ \cite {Axel3}. 

In the perpendicular direction, the superfluid fraction (\ref{sup1}) takes the form
\begin{equation}\label {supflui2}
\frac{n_s^{\perp}}{n} =1-2\pi R' \sqrt{\frac{n a^3}{\pi}} h^{\perp}(\epsilon_{dd}, \alpha)- \frac{\pi^2 \hbar}{45 mn c_s} \left(\frac{T}{\hbar c_s}\right)^4 {\cal Q}_{-5}^{\perp}(\epsilon_{dd}),
\end{equation}
where the function 
\begin{align} \label{funcP2}
h^{\perp}(\epsilon_{dd}, \alpha)=& \int_0^\pi d\theta \frac{ \sin\theta(1-\cos^2\theta) S(\alpha)}{\sqrt{1+\epsilon_{dd} (3\cos^2\theta-1)}} \nonumber\\
&=h(\epsilon_{dd}, \alpha)-h^{\parallel}(\epsilon_{dd}, \alpha), 
\end{align}
is increasing with $\epsilon_{dd}$ for fixed $\sigma/\xi$ as is displayed in Fig.\ref{disP}.b. 
And ${\cal Q}_j^{\perp}(\epsilon_{dd})= {\cal Q}_j(\epsilon_{dd})-{\cal Q}_j^{\parallel} (\epsilon_{dd})$.

Third terms in (\ref{supflui1}) and (\ref{supflui2}) which represent the thermal contribution of $n_s^{\perp}$ and $n_s^{\parallel}$, are calculated at low temperatures $T\ll ng$.
Whereas, at $T\gg ng$, there is copious evidence that both thermal terms of $n_s$ concide with the noncondensed density of an ideal Bose gas.
Furthermore, we read off from Eqs.(\ref{supflui1}) and (\ref{supflui2}) that for $\epsilon_{dd}\leq 0.5$, the thermal contribution of $n_s^{\perp}$ is smaller than that of $n_s^{\parallel}$, 
while the situations is inverted for  $\epsilon_{dd}> 0.5$.

For $\sigma/\xi\rightarrow 0$ and $\epsilon_{dd}=0$,  both components of the superfluid fraction reduce to $n_s/n=1-4n_R/3n$,
which well recoves earlier results of Refs \cite {Huang, Gior, Lopa} for isotropic contact interaction. 
For $\sigma/\xi\rightarrow 0$, we have $h^{\parallel}(\epsilon_{dd}, 0) ={\cal Q}^{\parallel}_{-1}(\epsilon_{dd})$ and $h^{\perp}(\epsilon_{dd}, 0) ={\cal Q}^{\perp}_{-1}(\epsilon_{dd})$. 
As a result, the disorder correction to superfluid fraction (\ref{depdis1}) becomes identical to that obtained in 3D dipolar BEC with delta correlated disorder \cite {Axel3}. 

We should stress also that for increasing $\varepsilon_{dd}$,  $h^{\parallel}(\epsilon_{dd}, \alpha)$ decreases, whereas $h(\epsilon_{dd}, \alpha)$ 
increases for fixed $\sigma/\xi$.
Therefore, this reveals that there exists a critical value of interaction $\epsilon^c_{dd}$ beyond which the system has the surprising property that
the disorder-induced depletion of the parallel superfluid density is smaller than the condensate depletion even at $T=0$.
This can be attributed to the fact that the localized particles can not contribute to superfluidity and, hence, form
obstacles for the superfluid flow. 
For large disorder correlation length i.e. $\sigma \gg \xi$, $\epsilon^c_{dd}$ reduces indicating that the localized particles are
localized in the respective minima for the disorder potential only for a finite localization time \cite {Axel4}.
This localization time remains to be analyzed in more detail in a future work.
In addition,  the superfluid fraction can be either larger or smaller than the condensate fraction $n_c/n=1-\tilde{n}/n$, depending on temperature, on interaction 
and on the strength of disorder. Increasing $R'$ leads to the simultaneous disappearance of the superfluid and condensate fractions. 

Note that the sound velocity  of a dipolar BEC in a weak external disorder potential can be calculated within the hydrodynamic approach as 
$c_s^2({\bf q})=(\partial \mu/m\partial n) {\bf q}^T \hat n_s {\bf q}$ \cite{Axel1, Axel2},
where the tensorial property of the superfluid density has been taken into account. 
From Eqs.(\ref{supflui1}) and (\ref{supflui2}) it follows that the sound velocity can also be separated into a parallel and a perpendicular components.
Both components change via effects of the interaction strength $\epsilon_{dd}$, disorder strength $R'$ and the ratio $\sigma/\xi$.
One can easily show also that the sound veclocity is consitent with the inverse compressibility $\kappa^{-1}=n^2\partial \mu/\partial n$\cite {Lev}, where  
the increase in $\kappa^{-1}$ tends to increase the sound velocity and vice versa.

\section{Conclusion}\label{conc}

In this paper, we have studied the properties of a homogeneous dipolar Bose gas in the presence of a weak disorder with 
autocorrelation function for an isotropic 3D laser speckle potential  at finite temperature. 
Using the Bogoliubov approach, we have calculated the condensate fluctuation due to disorder, 
as well as the corresponding corrections the condensed depletion, the anomalous fraction, the chemical potential, the ground state energy and the sound velocity. 
We have pointed out that the interplay between the anisotropy of DDI and the external random potential leads to modify both the BEC and the superfluidity characteristics.
Furthermore, we have reproduced the expression of the condensate fluctuations and thermodymanics quantities
obtained in the literature in the absence of both the DDI and the disordered potential. 
We discuss the validity criterion of the Bogoliubov approach in a dirty dipolar BEC.

Finally, an interesting question that begs to be asked is how the interplay of disorder and DDI can affect Anderson localization, 
or the quantum phases that arise due to disorder in the regime of strong correlations.


\end{document}